\documentclass{statsoc}

\usepackage{amsfonts,amssymb,amsmath,amscd,latexsym,color}
\usepackage{natbib}
\bibpunct{(}{)}{,}{a}{,}{,}

\usepackage{graphicx}
\usepackage{fullpage}

\newtheorem{example}{{\bfseries Example}}

\def \be{\begin{equation}}
\def \ee{\end{equation}} 

\def\1{1\!{\rm l}}  

\title[Discussion about the sBIC criterion]{Some comments about A Bayesian criterion for singular models}

\author{Christian P.~Robert}
\address{Universit{\' e} Paris-Dauphine, PSL Research University, CNRS, CEREMADE, 75775 Paris cedex 16, France; 
CREST, Paris, France; and  Department of Statistics, University of Warwick, Coventry, UK}
\email{xian@ceremade.dauphine.fr,robert@ensae.fr}
\author[Rousseau {\it et al.}]{Judith Rousseau}
\address{Universit{\' e} Paris-Dauphine, PSL Research University, CNRS, CEREMADE, 75775 Paris cedex 16, France; and CREST, Paris, France.}
\email{rousseau@ensae.fr}

\begin{document}

\maketitle




It is a well-known fact that the BIC approximation of the marginal likelihood in a given irregular  model $\mathcal M_k$
fails or may fail.  The BIC approximation has the form 
$$
BIC_k = \log p(\mathbf Y_n| \hat \pi_k, \mathcal M_k) - d_k \log n /2
$$ where $d_k $ corresponds on the number of parameters to be estimated in model $\mathcal M_k$. In irregular models
the dimension $d_k$ typically does not provide a good measure of complexity for model $\mathcal M_k$, at least in the
sense that it does not  lead to an approximation of
$$
\log m(\mathbf Y_n |\mathcal
M_k) = \log \left( \int_{\mathcal M_k}  p(\mathbf Y_n| \pi_k, \mathcal M_k) dP(\pi_k|k )\right) \,.
$$ 
A way to understand the behaviour of$\log m(\mathbf Y_n |\mathcal M_k) $ is through the \textit{effective dimension} 
$$ 
\tilde d_k  = -\lim_n \frac{  \log P( \{ KL(p(\mathbf Y_n| \pi_0, \mathcal M_k) ,
p(\mathbf Y_n| \pi_k, \mathcal M_k) ) \leq 1/n | k ) }{ \log n} 
$$ 
when it exists, see for instance the discussions in \citet{chambaz:rousseau:2008} and \cite{rousseau:07}. 

 \citet{watanabe:09} provided a more precise formula, which is the starting point of the approach of Drton and 
Plummer: 
$$
\log m(\mathbf Y_n |\mathcal M_k) = \log p(\mathbf Y_n| \hat \pi_k,
\mathcal M_k) - \lambda_k(\pi_0) \log n + [m_k(\pi_0) - 1] \log \log n + O_p(1) 
$$
where $\pi_0$ is the true parameter.

 The authors propose a clever algorithm to approximate of the marginal likelihood. 
 
 Given the popularity of the BIC criterion for model choice, obtaining a relevant penalized likelihood when the models
are singular is an important issue and we congratulate the authors for it. Indeed a major  advantage of the BIC formula
is that it is an off-the-shelf crierion which is implemented in many softwares, thus can be used easily by non
statisticians. 
 
 In the context of singular models, a more refined approach needs to be considered and although the algorithm proposed
by the authors remains quite simple, it requires that the functions $ \lambda_k(\pi)$ and $m_k(\pi)$ need be known
in advance, which so far limitates the number of problems that can be thus processed. 
In this regard their equation (3.2) is both puzzling and attractive. Attractive because it invokes nonparametric
principles to estimate the underlying distribution;  puzzling because why should we engage into deriving an approximation
like (3.1) and call for Bayesian principles when (3.1) is at best an approximation. In this case why not just use a true
marginal likelihood?

\section{ Why do we want to use a BIC type formula? } 

The BIC formula can be  viewed from  a purely frequentist perspective, as an example of  penalized likelihood. The
difficulty then stands into choosing the penalty and a common view on these approaches is to choose the smallest
possible penalty that still leads to consistency of the model choice procedure, isince it then enjoys better separation rates. In this case a $\log \log n$ penalty is
sufficient, as proved in \cite{gassiat:vanhandel:13}.

 Now whether or not this is a desirable property is entirely debatable, and one might advocate that  for a given sample size, if
the data fits the smallest model (almost) equally well, then this model should be chosen. But  unless one is specifying
what \textit{equally well} means, it does not add much to the debate. 
 
 This  also explains the popularity of the BIC formula (in regular models), since it
approximates the marginal likelihood and thus  benefits from the Bayesian justification of the measure of fit of a model
for a given data set, often qualified of being a Bayesian Ockham's razor.  But then why should we
not compute instead the marginal likelihood?
 
Typical answers to this question that are in favour of  BIC-type formula include:  (1) BIC is supposingly easier to
compute and (2) BIC does not call for a specification of the prior on the parameters within each model. Given that the
latter is a difficult task and that the prior can be highly influential in non-regular models, this may sound like a
good argument.  However, it is only apparently so, since the only justification of BIC is purely asymptotic, namely, in
such a regime the difficulties linked to the choice of the prior disappear. 
 
This is even more the case for the sBIC criterion,  since it is only valid if
the parameter space is compact. Then the impact of the prior becomes less of an issue as non informative priors can typically be used.  
With all due respect, the solution proposed by the authors, namely to use the posterior mean or the posterior mode to
allow for non compact parameter spaces, does not seem to make sense in this regard since they depend on the prior. The same comments apply
to the author's discussion on \textit{Prior's matter for sBIC}. Indeed variations of the sBIC could be obtained by
penalizing for bigger models via the prior on the weights, for instance as in \cite{mengersen:rousseau:2011} or by,
considering repulsive priors as in \cite{petralia:rao:dunson12}, but then it becomes more meaningful to (again) directly
compute the marginal likelihood. 

Remains (as an argument in its favour) the relative computational ease of use of sBIC, when compared with the marginal likelihood. 
This simplification is however achieved at the expense of requiring a deeper knowledge on the behaviour of the models
and it therefore looses the off-the-shelf appeal of the BIC formula and the range of applications of the method, at least so far. 

Although  the dependence of the approximation of $\log m(\mathbf Y_n |\mathcal M_k)$  on
$\mathcal M_j $, $j \leq k$  is strange, this does not seem crucial, since marginal likelihoods in themselves bring little information and they are only meaningful when compared to other marginal likelihoods. It becomes much more of an issue in the context of a large number of models. 

\section{ Should we care so much about penalized or marginal likelihoods ? } 

Marginal or penalized likelihoods are exploratory tools in a statistical analysis, as one is trying to define a reasonable model to fit the data. An unpleasant feature of these tools is that they provide numbers which in themselves do not have much meaning and can only be used in comparison with others and without any notion of uncertainty attached to them. 

A somewhat richer approach of exploratory analysis is to \textit{interrogate} the posterior distributions by either
varying the priors or by varying the loss functions. The former has been proposed in
\cite{vhavre:white:mengersen:rousseau15} in mixture models using the prior tempering algorithm. The latter has been used for instance by \cite{yau:holmes:13} for
segmentation based on Hidden Markov models. Introducing a decision-analytic perspective in the construction of
information criteria sounds to us like a reasonable requirement, especially when accounting for the current surge in
studies of such aspects.

\section*{Acknowledgements}
A preliminary version of these comments was written while the first author was visiting Monash University, Melbourne,
Australia. He gratefully acknowledges the support of Australian Research Council Discovery Grant No. DP15010172 towards
this visit and the congenial environment of the Department of Econometrics and Business Statistics at Monash. The
research of the first author is also supported by an IUF Senior Chair, 2016-2021.


\begin{thebibliography}{}

\bibitem[\protect\citeauthoryear{Chambaz and Rousseau}{Chambaz and
  Rousseau}{2008}]{chambaz:rousseau:2008}
Chambaz, A. and J.~Rousseau (2008).
\newblock Bounds for {B}ayesian order identification with application to
  mixtures.
\newblock {\em Ann. Statist.\/}~{\em 36}, 938--962.

\bibitem[\protect\citeauthoryear{Gassiat and van Handel}{Gassiat and van
  Handel}{2013}]{gassiat:vanhandel:13}
Gassiat, E. and R.~van Handel (2013).
\newblock Consistent order estimation and minimal penalties.
\newblock {\em IEEE Trans. Inform. Theory\/}~{\em 59}, 1115Ð1128.

\bibitem[\protect\citeauthoryear{Petralia, Rao, and Dunson}{Petralia
  et~al.}{2012}]{petralia:rao:dunson12}
Petralia, F., V.~Rao, and D.~B. Dunson (2012).
\newblock Repulsive mixtures.
\newblock In F.~Pereira, C.~J.~C. Burges, L.~Bottou, and K.~Q. Weinberger
  (Eds.), {\em Advances in Neural Information Processing Systems 25}, pp.\
  1889--1897. Curran Associates, Inc.

\bibitem[\protect\citeauthoryear{Rousseau}{Rousseau}{2007}]{rousseau:07}
Rousseau, J. (2007).
\newblock Approximating interval hypotheses: p-values and {B}ayes factors.
\newblock In J.~M. Bernardo, M.~Bayarri, J.~O. Berger, A.~P. Dawid,
  D.~Heckerman, A.~F.~M. Smith, and M.~West (Eds.), {\em Bayesian Statistics
  8}. Oxford: Oxford University Press.

\bibitem[\protect\citeauthoryear{Rousseau and Mengersen}{Rousseau and
  Mengersen}{2011}]{mengersen:rousseau:2011}
Rousseau, J. and K.~Mengersen (2011).
\newblock Asymptotic behaviour of the posterior distribution in overfitted
  mixture models.
\newblock {\em J. Royal Statist. Society Series B\/}~{\em 73\/}(5), 689--710.

\bibitem[\protect\citeauthoryear{van Havre, White, Mengersen, and Rousseau}{van
  Havre et~al.}{2015}]{vhavre:white:mengersen:rousseau15}
van Havre, Z., N.~White, K.~Mengersen, and J.~Rousseau (2015).
\newblock Overfitting bayesian mixture models with an unknown number of
  components.
\newblock {\em PLoS ONE\/}~{\em 10}.

\bibitem[\protect\citeauthoryear{Watanabe}{Watanabe}{2009}]{watanabe:09}
Watanabe, S. (2009).
\newblock Algebraic geometry and statistical learning theory.
\newblock In {\em Cambridge {M}onographs on {A}pplied and {C}omputational
  {M}athematics}, Volume~25. Cambridge University Press.

\bibitem[\protect\citeauthoryear{Yau and Holmes}{Yau and
  Holmes}{2013}]{yau:holmes:13}
Yau, C. and C.~C. Holmes (2013).
\newblock A decision-theoretic approach for segmental classification.
\newblock {\em Annals of Applied Statistics\/}~{\em 7}, 1814--1835.

\end{thebibliography}
\hyphenation{Post-Script Sprin-ger}

\end{document}